# Double Gate Bias Dependency of Low Temperature Conductivity of SiO$_2$-Si-SiO$_2$ Quantum Wells

M. Prunnila, J.M. Kivioja and J. Ahopelto

*VTT Micro and Nanoelectronics, Tietotie 3, Espoo, P.O. Box 1000, FI-02044 VTT, Finland*

**Abstract.** The gate bias dependency of conductivity σ is examined in two Si quantum wells with well thickness $t_w$ = 7 nm and $t_w$ = 14 nm. The conductivity of the thinner device behaves smoothly whereas the thicker device shows strong non-monotonic features as a function of gate voltages. We show that a strong minimum in σ occurs close to the threshold of second sub-band. Another minimum is seen at high electron density at symmetric well potential. This feature is addressed to sub-band wave function delocalization in the quantization direction.

**Keywords:** two-dimensional electron gas, bi-layer, resonant coupling, electron mobility, silicon
**PACS:** 73.40.-c, 72.20.-i

## INTRODUCTION

Development of silicon-on-insulator technology has enabled fabrication of silicon heterostructures where thin single crystalline Si film is sandwiched between amorphous SiO2 layers. This kind of SiO2-Si-SiO2 quantum well provides a unique material system where electron density can be tuned in a broad range due to high potential barriers formed by the SiO2 layers. In this work we examine low temperature conductivity of double gated SiO2-Si-SiO2 quantum wells.

## EXPERIMENTAL

The samples were fabricated on commercially available (100) silicon-on-insulator wafer as described in detail in Ref. [1]. The cross-sectional sample structure is illustrated in Fig. 1. Here we focus on two samples: F1 with Si well thicness $t_w$ = 7 nm and F2 with $t_W$ = 14 nm. The low temperature peak mobilities of F1 and F2 are ~ 7800 and ~ 14000 cm$^2$/Vs, respectively.

## RESULTS AND DISCUSSION

Figures 2(a) and (b) show conductivity σ as a function of top ($V_{TG}$) and back gate voltages ($V_{BG}$) measured from devices F1 and F2 at 270 mK, respectively. Fig. (b) also shows the electron density $n$ along the axis. Conductivity of both samples behaves qualitatively similarly when $V_{TG} \leq 0$ or $V_{BG} \leq 0$. The asymmetry in σ with respect to balanced gate line $V_{BG} = V_{TG} t_{BOX}/t_{OX}$ in both samples arises from different disorder (surface roughness) of the Si-BOX and Si-OX interfaces [1]. Note that sample F1 has lower mobility due to thinner well, which enhances surface roughness scattering.

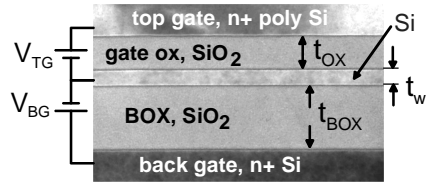

**FIGURE 1.** TEM image of double gate Si/SiO2 quantum well structure ($t_w$ = 18 nm) and illustration of gate biasing in the experiments.

In the bias window $V_{TG} \leq 0$ or $V_{BG} \leq 0$ the density dependency of mobility and conductivity of both devices exhibit the typical features of field-effect transistors [1-3]. However, when $V_{TG,BG} \geq 0$ this is no longer the case for F2 and σ shows strong non-monotonic features. The carrier density regime in Figs. 2(a) and 2(b) is $n$ ~ 0 - 4.0x10$^{16}$ m$^{-2}$ and F1 has only single sub-band occupied in this density regime [1]. Whereas, F2 has single and two sub-band gate bias windows depending on the gate balance. These different windows can be detected from Fig. 2(c), which shows a gray scale plot of normalized diagonal resistivity $\rho_{xx}(B)/\rho_{xx}(B=0)$ of F2 at magnetic field of B = 2.5 T. Two Landau level (LL) filling factors ν = nh/eB = 12, 20 are also identified in the Figure. In the single sub-band regimes the $\rho_{xx}$ minima, related to particular LL filling, are continuous trajectories. When two sub-bands are occupied the trajectories are broken into a 2D interference-like pattern. Note that the $\rho_{xx}$ minima are weaker when $V_{BG} > V_{TG}t_{BOX}/t_{OX}$ due to larger disorder of the Si-BOX interface.

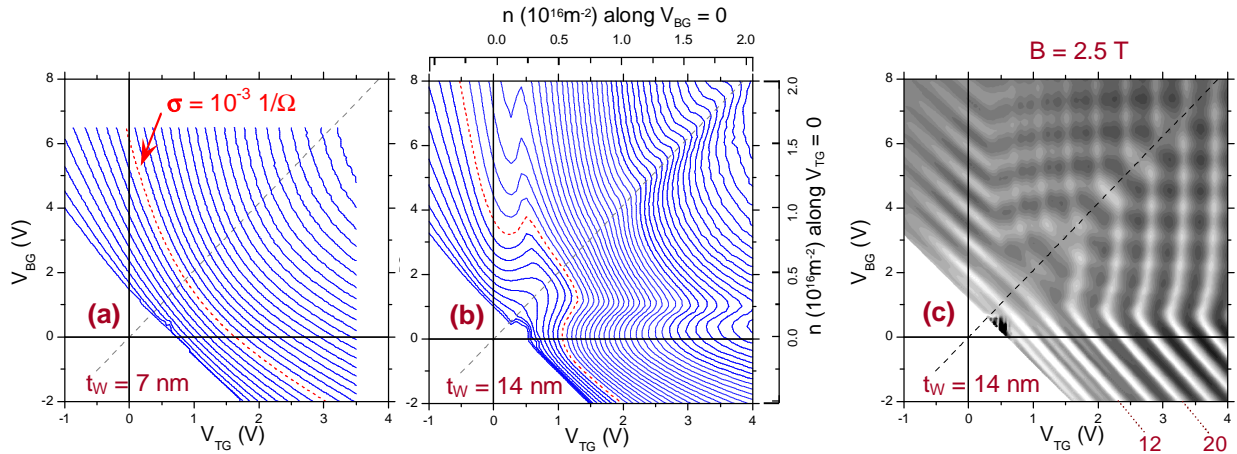

**FIGURE 2.** Contour plot of conductivity as a function of top and back gate voltages of (a) F1 (tw = 7 nm) and (b) F2 (tw = 14 nm). The contour spacing is $10^{-4}$ 1/Ω and the dashed curve is σ = $10^{-4}$ 1/Ω contour. (c) Normalized diagonal resistivity $\rho_{xx}(B)/\rho_{xx}(B=0)$ at B=2.5 T. Dark and light color correspond to high and low $\rho_{xx}$, respectively. All data measured at 270 mK.

Let us now further study the behavior of F2 at relatively high electron densities as a function of $V_{BG}$ at constant $V_{TG}$. We choose $\sigma(V_{TG}=const., V_{BG})$ $V_{TG} \geq$ 2.0 V. Now starting from $V_{BG}$ = -2.0 V and moving towards positive $V_{BG}$ we can observe that first there is a maximum in $\sigma(V_{TG}=const., V_{BG})$ at $V_{BG} \sim 0$ which is followed by a minimum at $V_{BG} \sim 1.0$ V (see Ref.[3] for the actual curves). From Fig. 2(c) we see that the minimum occurs in the vicinity of the 2$^{nd}$ sub-band threshold. (This feature is also seen in the Hall effect [2].) Similar effect has been observed in compound semiconductor double quantum wells (bi-layers) where conductivity decreases at the threshold of the second well [4]. This behavior has been addressed to interlayer scattering. Note that the sub-band wave functions are localized at the opposite interfaces in device F2 [3] especially at high electron densities. Thus, F2 resembles a bi-layer system, which is of course typical for wide quantum wells. The observed minimum in $\sigma(V_{TG}=const., V_{BG})$ a at $V_{BG} \sim 1.0$ V, therefore, most likely originates from inter-layer or inter-sub-band effects. It is interesting to note that a recent study reports on a step in the conductivity at the boundary of (zero magnetic field) valley-unpolarized and -polarized regions in high disorder Si-well, which can be recognized also as an inter-layer type of effect [5].

If $V_{BG}$ is increased further we observe another minimum, which occurs at balanced gate line. At balanced gate bias the well potential is symmetric and the sub-band wave functions delocalize across the well, which increases scattering and leads to drop in the conductivity [3]. This type of effect is referred as resistance resonance [6].

There is still one feature in σ that is different in F1 and F2: if we move along the ν = 8 - 12 trajectories, which correspond to $n = 4.8 - 7.3 \times 10^{15}$ m$^{-2}$ we can see that there is a local minimum in σ of F2 at balanced gate line. Note that this is single sub-band regime for F2 ($\rho_{xx}$ trajectories are continuous, see also [3]), yet F1 shows no such feature. In Ref. [3] it is speculated that this behavior in F2 arises from similar effect as the minimum in $\sigma(V_{TG}=const., V_{BG})$ at $V_{BG} \sim 1.0$ V. Note, however, that in density range $n \sim 5 - 10 \times 10^{15}$ m$^{-2}$ Coulomb and surface roughness scattering have almost equal strength. This complicates the situation and numerical calculations of scattering rates are required to understand the overall behavior at low densities in the vicinity of the balanced gate bias.

## ACKNOWLEDGMENTS

M. Markkanen is thanked for assistance in the sample fabrication. Academy of Finland is acknowledged for financial support through project # 205467.